# The Radio Galaxy Environment Reference Survey (RAGERS): a submillimetre study of the environments of massive radio-quiet galaxies at z = 1–3

Thomas M. Cornish ,[1,2]★ Julie L. Wardlow ,[1]★ Thomas R. Greve ,[3] Scott Chapman ,[4,5] Chian-Chou Chen ,[6] Helmut Dannerbauer ,[7,8] Tomotsugu Goto ,[9] Bitten Gullberg ,[3] Luis C. Ho ,[10,11] Xue-Jian Jiang ,[12,13] Claudia Lagos ,[3,14,15] Minju Lee ,[3] Stephen Serjeant ,[16] Hyunjin Shim ,[17] Daniel J. B. Smith ,[18] Aswin Vijayan ,[3,19] Jeff Wagg[20]† and Dazhi Zhou [3,5]

*Affiliations are listed at the end of the paper*

**ABSTRACT**
Measuring the environments of massive galaxies at high redshift is crucial to understanding galaxy evolution and the conditions that gave rise to the distribution of matter we see in the Universe today. While high-$z$ radio galaxies (H$z$RGs) and quasars tend to reside in protocluster-like systems, the environments of their radio-quiet counterparts are relatively unexplored, particularly in the submillimetre, which traces dust-obscured star formation. In this study, we search for 850 μm-selected submillimetre galaxies (SMGs) in the environments of massive ($M_\star > 10^{11}$ M$_\odot$), radio-quiet ($L_{500\mathrm{MHz}} \lesssim 10^{25}$ WHz$^{-1}$) galaxies at $z \sim$ 1–3 using data from the SCUBA-2 COSMOS (S2COSMOS) survey. By constructing number counts in circular regions of radius 1–6 arcmin and comparing with blank-field measurements, we find no significant overdensities of SMGs around massive radio-quiet galaxies at any of these scales, despite being sensitive down to overdensities of $\delta \sim$ 0.4. To probe deeper than the catalogue we also examine the distribution of peaks in the SCUBA-2 signal-to-noise (SNR) map, which reveals only tentative signs of any difference in the SMG densities of the radio-quiet galaxy environments compared to the blank field, and only on smaller scales (1 arcmin radii, corresponding to ∼ 0.5 Mpc) and higher SNR thresholds. We conclude that massive, radio-quiet galaxies at cosmic noon are typically in environments with $\delta \lesssim 0.4$, which are either consistent with the blank field or contain only weak overdensities spanning sub-Mpc scales. The contrast between our results and studies of H$z$RGs with similar stellar masses and redshifts implies an intrinsic link between the wide-field environment and the radio luminosity of the active galactic nucleus at high redshift.

## 1 INTRODUCTION

A link between galaxies and their environments has long been observed in the local Universe, with local elliptical galaxies predominantly residing in denser environments, such as galaxy clusters, while spirals are generally found in the field (Dressler 1980). This link is less clearly defined at high redshift, particularly at $z \gtrsim 1$ where galaxy clusters have not had sufficient time to virialize, though observations suggest that 25–50 per cent of massive galaxies up to $z \sim$ 1.5 lie in groups or low-mass clusters (e.g. Gerke et al. 2012; Knobel et al. 2012; Tempel et al. 2016; Boselli, Fossati & Sun 2022). Quasars and high-redshift radio galaxies (H$z$RGs) – massive ($M_\star \gtrsim 10^{10.5}$ M$_\odot$; Seymour et al. 2007) galaxies hosting radio-loud (rest-frame $L_{500\mathrm{MHz}}$ > $10^{27}$ W Hz$^{-1}$; Miley & De Breuck 2008) active galactic nuclei (AGN) – are often used as beacons for high-density environments at these epochs (e.g. Kurk et al. 2000; Venemans et al. 2002; Kuiper et al. 2011; Hayashi et al. 2012; Wylezalek et al. 2013; Cooke et al. 2014; Hall et al. 2018). Furthermore, H$z$RGs have other properties consistent with the progenitors of central brightest cluster galaxies (BCGs; e.g. Best, Longair & Rottgering 1997; Miley & De Breuck 2008): in addition to their high stellar masses, they simultaneously display signatures of high star formation activity (e.g. Dey et al. 1997; Villar-Martín et al. 2007) and old stellar populations (e.g. Best, Longair & Roettgering 1998; Jarvis et al. 2001; Rocca-Volmerange et al. 2004) indicative of complex star formation histories, as expected of the progenitors of BCGs.

However, whilst it is well-established that H$z$RGs and quasars trace galaxy overdensities, it is currently debated whether this is due to their typically high stellar masses, or whether interactions between the AGN activity and local environment leads to these observations (e.g. see discussion in Wylezalek et al. 2013). If the high masses of H$z$RGs are the driver behind their observed environment then one would expect to find radio-quiet (RQ) galaxies with similar masses inhabiting similarly overdense environments, yet observational evidence suggests this may not be the case. For example, Hatch et al. (2014) used galaxies selected with the *Spitzer*/Infrared Array Camera (IRAC) to compare the environments of radio-loud AGN (including a sample of 208 H$z$RGs) from the CARLA survey (Wylezalek et al. 2013) with those of a radio-quiet control sample matched in stellar mass and redshift, finding that galaxies in the control sample reside in significantly less dense environments. This is further supported by simulations, in which H$z$RGs are seen to be hosted by more massive dark matter haloes than RQ galaxies with the same stellar mass, due to AGN feedback preventing the build-up of stellar mass in the H$z$RGs (e.g. Izquierdo-Villalba et al. 2018).

★ E-mail: thomas.cornish@physics.ox.ac.uk (TMC); j.wardlow@lancaster.ac.uk (JLW)
† PIFI Visiting Scientist, Purple Mountain Observatory, Np. 8 Yuanhua Road, Qixia District, Nanjing 210034, People's Republic of China.

Studies of galaxy environments at high redshift have predominantly involved identifying overdensities of galaxies selected in the rest-frame ultraviolet (UV) to optical, e.g. via Ly$\alpha$ or H$\alpha$ emission (e.g. Kurk et al. 2000; Venemans et al. 2007; Shimakawa et al. 2018), or through the Lyman or Balmer/4000 Å breaks (e.g. Wold et al. 2000, 2001; Kajisawa et al. 2006; Hatch et al. 2011; Uchiyama et al. 2022). Such studies implicitly omit the population of dusty star-forming galaxies (DSFGs) now known to be a significant contributor to the total star-formation rate density at these redshifts (e.g. Coppin et al. 2006; Barger et al. 2012; Swinbank et al. 2014). These DSFGs are highly obscured in the rest-frame UV-to-optical due to the high abundance of dust, instead being most luminous in the far-infrared (FIR)/submillimetre (with the brightest of these galaxies being labelled 'submillimetre galaxies', or SMGs). A complete picture of protocluster formation thus also requires dedicated FIR/submillimetre studies. For a review of cluster evolution in the infrared see Alberts & Noble (2022).

FIR and submillimetre observations of known high-redshift protoclusters have confirmed overdensities of SMGs in these structures (e.g. Blain et al. 2004; Tamura et al. 2009; Matsuda et al. 2011), and simulations confirm that SMGs can trace overdensities (e.g. Davé et al. 2010). However, there is still debate as to whether SMGs as a population typically reside in protoclusters, or whether they may simply be effective as tracers of such structures (e.g. Miller et al. 2015; Casey 2016; Calvi, Castignani & Dannerbauer 2023; Cornish et al. 2024). FIR and submillimetre surveys targeting H$z$RGs have identified dust-obscured emission for the H$z$RGs themselves, as well as identifying nearby populations of SMGs, in excess of the numbers expected in the field (e.g. Ivison et al. 2000; Stevens et al. 2003, 2010; Greve et al. 2007; Dannerbauer et al. 2014; Rigby et al. 2014; Zeballos et al. 2018). Targeted searches for radio galaxies associated with known overdensities of SMGs complement this picture detecting radio sources whose emission significantly exceeds expectations from star formation alone, suggesting the presence of a radio-loud AGN (e.g. Oteo et al. 2018; Chapman et al. 2023). SMG populations have also been observed around high-redshift quasars with a range of properties (e.g. Decarli et al. 2017; Fan et al. 2017; Li et al. 2020; Wethers et al. 2020; Bischetti et al. 2021; García-Vergara et al. 2022; Nowotka et al. 2022; Arrigoni Battaia et al. 2023; Li et al. 2023), however not all of these populations are found to be overdense relative to the blank field, and the relative importance of the quasar radio luminosity is either not considered or poorly constrained due to small sample sizes.

There are currently few studies of the FIR and submillimetre environments of high-mass and high-redshift, but radio-quiet galaxies. Rigby et al. (2014) found a hint of a correlation between radio power and the overdensity of surrounding *Herschel*-detected galaxies, though they only probe down to radio luminosities of $L_{500\mathrm{MHz}} \sim 10^{28.5}$ W. Nevertheless, the result from Rigby et al. (2014), along with similar low-significance studies at shorter wavelengths (e.g. Galametz et al. 2012), may indicate that RQ galaxies have fewer (submillimetre) companions than H$z$RGs, which suggests the potential for different evolutionary pathways in these populations.

The RAdio Galaxy Environment Reference Survey (RAGERS; Greve et al., in prep.) is a James Clark Maxwell Telescope (JCMT)/Submillimetre Common-User Bolometer Array 2 (SCUBA-2) Large Program (program ID: M20AL015) with the aim of mapping the submillimetre environments of 27 powerful H$z$RGs at $1 < z < 3.5$ and comparing them to the environments of a mass- and redshift-matched RQ control sample. In this paper, we use archival SCUBA-2 data to measure the environments of this control sample, and compare the SMG densities with blank field expectations to assess whether the environments of massive high-redshift RQ galaxies are overdense.

This paper is structured as follows: in Section 2, we describe our sample selection; Section 3 details the method used to measure the submillimetre environments; Section 4 and Section 5 contain our key results and subsequent discussion; we present our conclusions in Section 6. Throughout this paper we assume a flat Lambda cold dark matter ($\Lambda$CDM) cosmology with $\Omega_\mathrm{M} = 0.315$, $\Omega_\Lambda = 0.685$, and $H_0 = 67.4$ km s$^{-1}$ Mpc$^{-1}$ (Planck Collaboration 2020). Physical scales with this cosmology are 0.495 Mpc arcmin$^{-1}$ at $z = 1$ and 0.449 Mpc arcmin$^{-1}$ at $z = 3.5$.

## 2 DATA

The aim of this study is to measure the density of SMGs around a large sample of massive, RQ galaxies, where these RQ galaxies have similar redshifts and stellar masses to the radio-loud (RL) galaxies targeted by RAGERS. The 27 H$z$RGs in the RAGERS sample are approximately uniformly distributed at $z = 1$–3.5 and were selected from the *Herschel* Radio Galaxy Evolution Project (HeRGÉ; Seymour et al. 2007; De Breuck et al. 2010; Seymour et al. 2012). With stellar masses and radio luminosities in the ranges of $\log(M_\star/\mathrm{M}_\odot) \sim 11.0$–11.9 and $\log(L_{500\,\mathrm{MHz}}/\mathrm{W\,Hz}^{-1}) \sim 28.2$–29.2, respectively, the RAGERS targets are representative of typical H$z$RGs at $z = 1$–3.5 (De Breuck et al. 2010; Greve et al. in prep). Thus by selecting RQ analogues to the RAGERS galaxies our study is broadly applicable and representative of RQ galaxies in comparison to the whole H$z$RG population at these redshifts.

In order to select appropriate RQ galaxies and study their submillimetre environments, we require: (1) high-quality redshifts and stellar masses for all RQ galaxies for comparison with the RL sample; (2) RQ galaxies must reside in areas with contiguous, deep submillimetre coverage to enable identification of companion SMGs; (3) the targets must have been observed in the radio, to enable the exclusion of radio-loud galaxies from our study.

We therefore choose the Cosmic Evolution Survey (COSMOS) field as the location for our study, due to the wide area, deep radio and submillimetre data, and extensive optical and near-IR (NIR) photometry and spectroscopy. The COSMOS2020 catalogue (Weaver et al. 2022) contains extensive UV-to-NIR photometry covering the entire 2 deg$^2$ of the COSMOS field, as well as high-quality photometric redshifts and galaxy properties (such as stellar masses) derived from spectral energy distribution fitting. The Very Large Array (VLA)-COSMOS survey adds extensive radio coverage at 1.4 and 3 GHz (Schinnerer et al. 2004; Smolčić et al. 2017a), and SCUBA-2 submillimetre observations at 850 μm are provided by the SCUBA-2 COSMOS survey (S2COSMOS; Simpson et al. 2019, henceforth S19). S2COSMOS reaches a median noise level of $\sigma_{850\mu\mathrm{m}} = 1.2$ mJy beam$^{-1}$ over the deepest 1.6 deg$^2$ of the survey (the 'MAIN' region), though there is an additional 1 deg$^2$ 'SUPP' region with shallower data (median $\sigma_{850\mu m} = 1.7$ mJy beam$^{-1}$), which we exclude from our analyses. Combining the data from COSMOS2020, VLA-COSMOS, and S2COSMOS enables the effective identification of RQ analogues for the RAGERS H$z$RGs and measurement of their submillimetre environments.

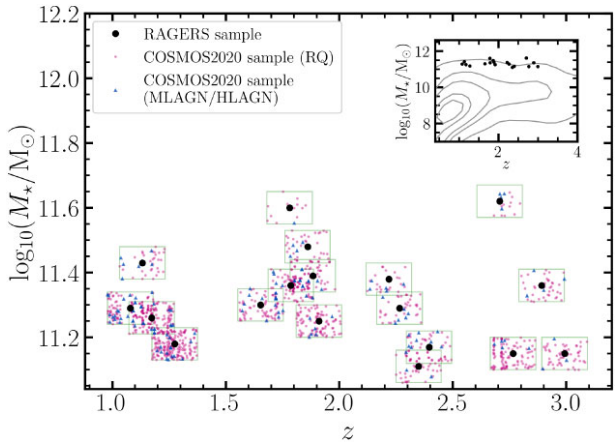

**Figure 1.** Distribution of the RAGERS radio-loud galaxies (black circles) in stellar mass and redshift, with green boxes to show the selection criteria for identifying radio-quiet analogues used in this study. Coloured markers show all galaxies from COSMOS2020 (Weaver et al. 2022) with redshifts and stellar masses that satisfy the selection criteria, and also reside in regions with a local rms noise of $\lesssim 1.3$ mJy beam$^{-1}$ in the S2COSMOS 850 µm map (Simpson et al. 2019). Blue triangles show the subset of these galaxies that are flagged as MLAGN or HLAGN in VLA-COSMOS (Smolčić et al. 2017b), while pink dots show galaxies which do not have this flag and thus make it into the final sample of RQ analogues. The inset panel shows the stellar masses and redshifts of RAGERS H$z$RGs overlaid on contours marking the 10th, 30th, 50th, 70th, and 90th percentiles of galaxies in the entire COSMOS2020 catalogue, demonstrating that H$z$RGs (and thus our RQ analogues) have high-stellar masses for their redshifts.

### 2.1 Sample selection

In order to study RQ galaxies that are analogues of the RAGERS H$z$RGs, we require a sample of RQ galaxies with similar stellar masses and redshifts to RAGERS sources, so that a controlled comparison can ultimately be made with the H$z$RGs (Greve et al. in prep.). We therefore begin by selecting galaxies from COSMOS2020 (Weaver et al. 2022) that have stellar masses and redshifts within $\Delta \log(M_\star/M_\odot) = \pm 0.05$ and $\Delta z = \pm 0.1$ of each RAGERS H$z$RG (Fig. 1). The sizes of the redshift and stellar mass intervals were chosen such that each RAGERS H$z$RG has at least 10 RQ analogues in the final sample. Of the 27 RAGERS RL galaxies, nine have only have upper limits on their stellar masses (due to AGN contamination in the photometry) and are therefore excluded, and the remainder of this study focuses on RQ analogues to the remaining 18 H$z$RGs in RAGERS. Note that this criterion removes the three highest-redshift H$z$RGs, such that the remaining galaxies reside at $z = 1$–3.

Having selected COSMOS2020 sources with similar stellar masses and redshifts to RAGERS galaxies, we next identify and remove candidate RL galaxies using VLA-COSMOS (Smolčić et al. 2017b), which contains all 3 GHz sources that are detected at >5$\sigma$ (median $\sigma = 2.3$ mJy beam$^{-1}$) and have counterparts at optical, near-infrared, mid-infrared (MIR), or X-ray wavelengths. Sources in VLA-COSMOS are flagged as either moderate-to-high radiative luminosity AGN (HLAGN), low-to-moderate radiative luminosity AGN (MLAGN), or star-forming galaxies (SFGs) according to their multiwavelength properties (see Smolčić et al. 2017b for details). Using a matching radius of 1 arcsec we cross-match the COSMOS2020 catalogue with the optical/NIR/MIR counterpart positions listed in the VLA-COSMOS catalogue and exclude any sources flagged as HLAGN or MLAGN from our sample. We also remove any sources that have multiple radio components and sources with a probability >20 per cent of being falsely matched to their optical/NIR/MIR counterpart in VLA-COSMOS.

Finally, we use the sensitivity map from S2COSMOS (Simpson et al. 2019) to select sources with the deepest 850 µm coverage, where the depth is $\sigma_{rms} < 1.3$ mJy beam$^{-1}$. Our study probes the environments of RQ galaxies out to 6 arcmin, so we therefore also require that the deep SCUBA-2 coverage extends at least 6 arcmin radius around each RQ galaxy, and we exclude sources that are close to the edge of the deep SCUBA-2 regions.

The final sample consists of the 1128 galaxies that remain after applying all of these selection criteria. Of the 1128 galaxies there are between 11 and 185 RQ analogues (median 51) for each of the 18 RAGERS H$z$RGs. The stellar masses and redshifts of this sample of RQ galaxies, compared with the H$z$RGs from RAGERS and the whole COSMOS2020 catalogue are shown in Fig. 1.

It is worth noting that while our selection criteria do not enforce any explicit cuts in radio luminosity, it is possible to verify that these galaxies are significantly less radio-loud than the RAGERS H$z$RGs (for which $L_{500\,MHz} \gtrsim 10^{28.2}$ W Hz$^{-1}$; De Breuck et al. 2010). Of the 1128 galaxies in our RQ sample, 164 have counterparts in the VLA-COSMOS catalogue (14.6 per cent; Smolčić et al. 2017b). For these 164 galaxies we use the rest-frame 1.4 and 3 GHz luminosities from VLA-COSMOS to calculate the radio spectral index[1] and extrapolate to estimate the rest-frame 500 MHz luminosity. The distribution of the resultant rest-frame 500 MHz luminosities has a median of $L_{500\,MHz} = 10^{24.1 \pm 0.5}$ W Hz$^{-1}$, with a maximum of $L_{500\,MHz} \sim 10^{25.2}$ W Hz$^{-1}$. The remaining 964 galaxies in our RQ sample (85.4 per cent) have no counterpart in the VLA-COSMOS catalogue, and can thus be safely assumed to also satisfy $L_{500\,MHz} \lesssim 10^{25.2}$ W Hz$^{-1}$. The galaxies in our sample thus have radio luminosities at least three orders of magnitude lower than those of the RAGERS H$z$RGs, ensuring a sufficient dichotomy between the two samples.

## 3 CALCULATING NUMBER COUNTS

In order to probe the environments of RQ galaxies we measure the 850 µm number counts in apertures around the target galaxies. Number counts quantify the surface density of sources as a function of their submillimetre flux density; as such they provide the most direct measure of source abundance and environment. Number counts are particularly useful for submillimetre data as they do not require cross matching to other wavelengths or obtaining redshifts, both of which are challenging and biased for SMGs due to the large beam sizes of single-dish surveys and the faintness of these dusty, high-redshift galaxies at optical and NIR wavelengths (e.g. see Casey, Narayanan & Cooray 2014 for a review). In this study, we measure both the differential (d$N$/d$S$ [mJy deg$^{-2}$]) and cumulative ($N(>S)$ [deg$^{-2}$]) number counts around the RQ galaxies and the blank field. We use both differential and cumulative counts because of the different strengths and weaknesses of the two measurements: differential counts have the advantage that the measurements in each bin are relatively independent of each other, but cumulative counts contain more sources in the fainter bins and therefore have smaller uncertainties.

---
[1]For galaxies where only 3 GHz luminosities were measured, Smolčić et al. (2017b) assume a spectral index $\alpha = -0.7$ to estimate the 1.4 GHz luminosities.

Constructing number counts requires an understanding of how the completeness of the survey varies with observed parameters. S19 used simulations to map the variation in completeness of S2COSMOS as a function of deboosted 850 μm flux density and local rms noise and we use this result for our completeness correction. Following S19, we create $10^4$ realizations of the S2COSMOS catalogue, where each version is generated by randomly drawing (deboosted) flux densities for each submillimetre source based on the uncertainties in the original S2COSMOS catalogue. Completeness corrections are then calculated for each randomly drawn flux density and the local rms of the source using the completeness function derived in S19.

Number counts are derived for the environments of the RQ galaxies selected in Section 2 by considering one RQ analogue for each RAGERS H$z$RG at a time. These galaxies are randomly chosen but with the proviso that the same galaxy cannot be selected as an analogue for multiple H$z$RGs, as would otherwise be possible for RL galaxies with similar stellar masses and redshifts (see Fig. 1). We then identify submillimetre sources in S2COSMOS that lie within apertures of radius $R$ centred on each RQ galaxy. To assess the scale of any overdensities and examine whether the choice of radius affects the results we construct four versions of the number counts using radii of 1, 2, 4, and 6 arcmin. These radii correspond to physical scales of ∼0.5–3 Mpc at the redshifts probed by this study.

For each of the $10^4$ realizations of the S2COSMOS catalogue, the selected submillimetre sources are binned by their randomly drawn deboosted flux densities and each source is weighted by the reciprocal of the completeness corresponding to its local rms noise and deboosted flux density. For the differential number counts, the weighted counts in each bin are divided by the product of the bin width ($\Delta S$) and the combined area of the apertures used to survey the RQ galaxy environments ($A_{\rm tot}$); for the cumulative number counts, each bin is just divided by $A_{\rm tot}$. Each aperture is treated independently – i.e. any overlap between apertures is ignored, and sources within the overlapping area are multiply counted – such that $A_{\rm tot}$ is simply the sum of each aperture's area.

The random selection of RQ galaxies is repeated 1000 times and number counts constructed for all $10^4$ realizations of the catalogue each time, such that each bin ultimately has a distribution of $18 \times 10^7$ values associated with it. The width of this distribution encapsulates the uncertainties from both the source flux densities and the stochasticity in the selection of RQ analogues for each iteration. The final bin heights are then taken to be the medians of these distributions and the bin uncertainties incorporate both the 16th–84th percentile ranges in the distributions and Poissonian uncertainties.

To interpret the submillimetre number counts around RQ galaxies we require a measure of the number counts in the blank field. Whilst blank-field 850 μm counts have been studied previously (e.g. Casey et al. 2013; Hsu et al. 2016; Geach et al. 2017; Simpson et al. 2019; Garratt et al. 2023) we regenerate them using the MAIN sample from S2COSMOS ($A_{\rm tot} = 1.6$ deg$^2$) and our method and binning to ensure a direct like-for-like comparison. In generating the blank-field counts all $10^4$ realizations of the catalogue are used and we have verified that our results are consistent with those from S19 and S19 showed that COSMOS is similarly dense at 850 μm as other blank fields.

## 4 RESULTS

### 4.1 The environments of massive RQ galaxies

The differential and cumulative number counts for the RQ galaxy environments we calculated as described in Section 3 and are presented in Fig. 2. Results are shown separately for the four different search radii of 1, 2, 4, and 6 arcmin, alongside the blank-field results constructed from the MAIN S2COSMOS sample. The combined number counts for all the RQ analogues are highlighted and it is these that we use for the remainder of our analyses. We also show the number counts around the RQ analogues of each RAGERS H$z$RG to demonstrate the scatter between fields, though the small numbers involved mean that uncertainties on these subsets are significant. When considering the whole sample, there is qualitatively no significant difference between the number counts of the blank field and the RQ environment, regardless of the spatial scale considered.

To quantitatively determine whether the environments of RQ galaxies have different submillimetre number counts we fit them with Schechter functions (Schechter 1976). Differential number counts are typically parametrized using Schechter functions of the form

$$\frac{dN}{dS} = \frac{N_0}{S_0}\left(\frac{S}{S_0}\right)^{-\gamma}\exp\left(-\frac{S}{S_0}\right) \quad (1)$$

where $N_0$ and $S_0$ determine surface density and flux density at the 'knee' of the Schechter function, respectively, and $\gamma$ is the slope of the function at the faint end. By integrating equation (1) the same parameters are used to define a function to fit the cumulative counts:

$$N(>S') = \int_{S'}^{\infty}\frac{dN}{dS}dS = N_0\Gamma\left(-\gamma+1, \frac{S'}{S_0}\right) \quad (2)$$

where $\Gamma$ represents the upper incomplete gamma function.

The best-fitting parameters for the functions described by equations (1) and (2) are measured for both the blank field and the RQ environment number counts using a Markov Chain Monte Carlo (MCMC) fitting procedure. Bins with flux density < 3 mJy are excluded from the fitting due to low completeness. The $S_0$, $N_0$, and $\gamma$ measured for each region are summarized in Fig. 3 (including the correlations between parameters) and Table 1, and the corresponding best-fits are shown on the number counts in Fig. 2.

As shown by Fig. 3, at all scales examined (radii of 1–6 arcmin) there is no significant difference between the best-fitting parameters in regions close to RQ galaxies and the blank field. The significant overlap, even at the 1$\sigma$ level, indicates that there is no significant overdensity of 850 μm-selected SMGs in the environments of RQ galaxies.

Due to the sizeable uncertainties in our number counts and in fitting three-parameter Schechter functions, we also test for significant differences in the $N_0$ parameter when $S_0$ and $\gamma$ are both fixed to the blank-field values, i.e. for $S_0 = 3.1$ (3.0) mJy and $\gamma = 1.6$ (1.5) for the differential (cumulative) number counts. Since the $N_0$ parameter scales the number density of submillimetre sources (i.e. the $y$-axis on Fig. 2) a value of $N_0$ significantly above the blank-field value would imply an overdense environment. However, even with the added constraints of a single parameter fit (and the resulting smaller uncertainties), we still find no significant difference between the blank field and the environments of the RQ analogues (see Table 1). These results are in contrast to the environments of H$z$RGs, which have been found to contain overdensities of submillimetre sources (e.g. Ivison et al. 2000; Stevens et al. 2003, 2010; Greve et al. 2007; Rigby et al. 2014).

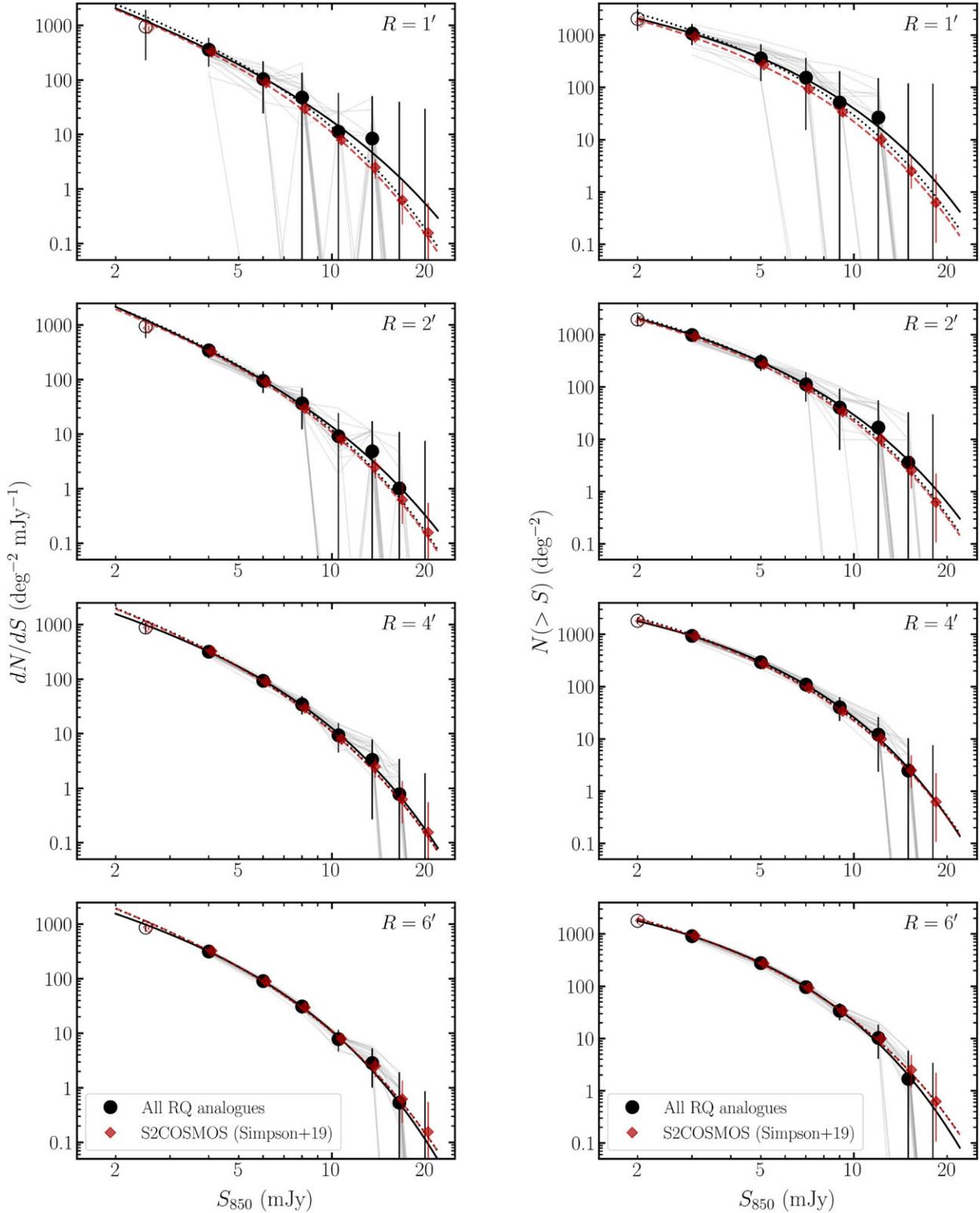

**Figure 2.** Differential *(left)* and cumulative *(right)* 850 μm number counts comparing the regions around RQ galaxies with the blank field. Each row shows the measurements and Schechter function fits using a different radius to search for candidate submillimetre companions (black circles and black lines), as indicated in the top-right corner of each panel (*top to bottom:* 1, 2, 4, 6 arcmin). Red diamonds and dashed lines represent the blank field and show the number counts and corresponding fits for the entire MAIN region of the S2COSMOS field, created using the method described in Section 3 and catalogue from S19. Faint, grey lines show the results for the RQ analogues of each individual RAGERS H$z$RG and give an indication of the scatter between different RQ galaxy regions (though the small number statistics means that uncertainties are significant). Solid black lines show the best-fitting Schechter functions for the combined data sets when all parameters are allowed to vary, and dotted black lines show the fits when $N_0$ is the only free parameter (i.e. $S_0$ and $\gamma$ are fixed to the blank field values; see Section 4.1). Bins with flux density < 3 mJy are marked with open symbols and excluded from the fitting due to low completeness. There is no significant difference between the submillimetre environments of the RQ sample and the blank field at any of these scales.

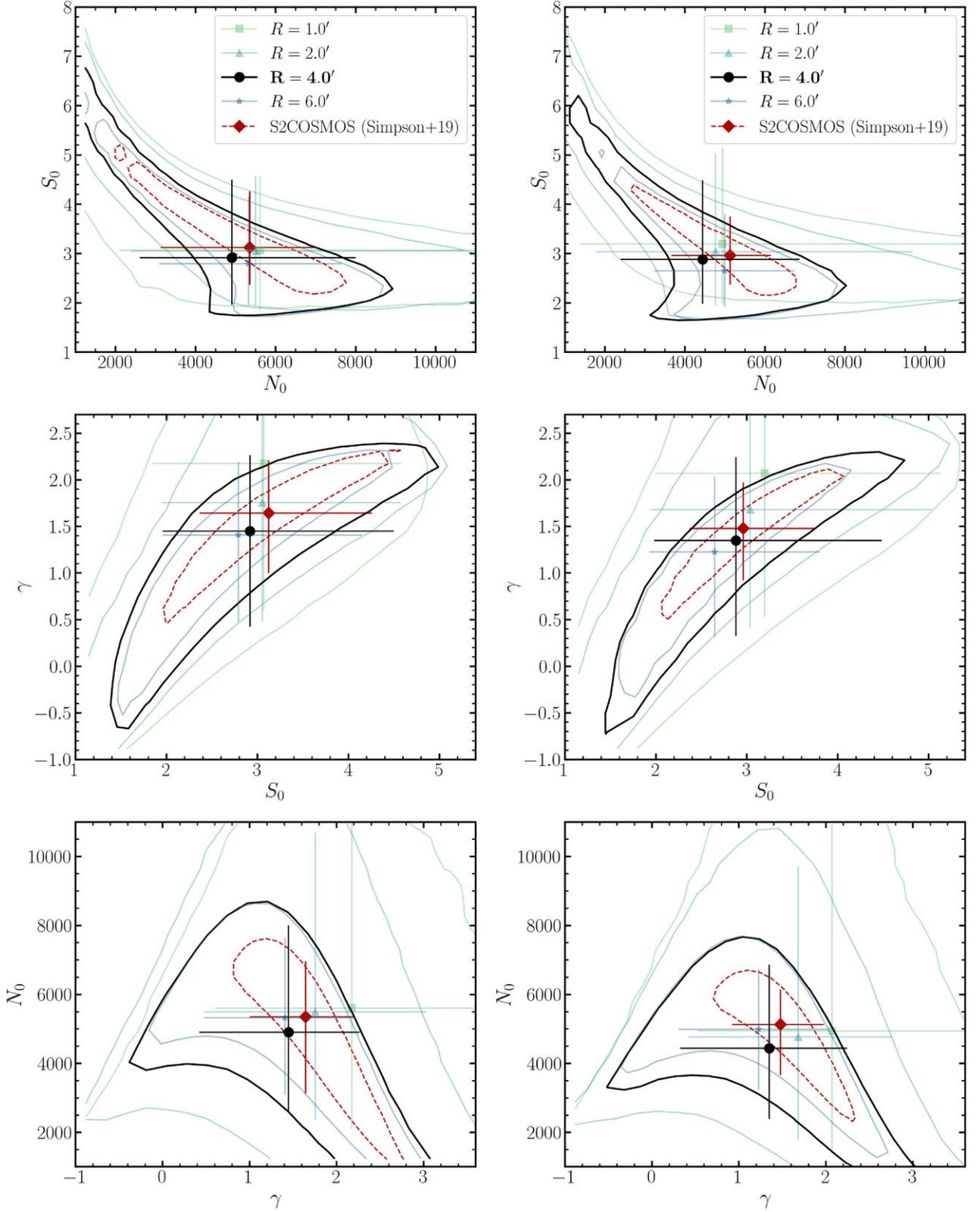

**Figure 3.** Contours representing the 1$\sigma$ confidence regions for the parameters of the Schechter functions fitted to the differential *(left)* and cumulative *(right)* number counts. Data points and error bars show the medians and 16th–84th percentile ranges on each individual parameter. Black circles and thick contours represent the fits to the RQ analogue number counts when a 4 arcmin radius is used; squares, upward triangles, and downward triangles are used for radii of 1, 2, and 6 arcmin, respectively. Red diamonds and dashed contours show the results for the blank field, based on the MAIN sample of S2COSMOS. There is significant overlap between the measurements at all radii and with the blank-field results, demonstrating that there is no significant difference between the 1–6 arcmin scale environments of our RQ analogues and the blank field.

**Table 1.** Best-fitting Schechter parameters for the differential and cumulative number counts, obtained from MCMC fitting. The values quoted are the medians of the posterior distributions for each parameter obtained through MCMC fitting, with $1\sigma$ uncertainties defined by the 16th and 84th percentiles. $N_0$ values in brackets show the results of fixing $S_0$ and $\gamma$ to the blank-field values.

|  | Radius (arcmin) | $N_0$ (deg$^{-2}$) | $S_0$ (mJy) | $\gamma$ |
|---|---|---|---|---|
| Differential | 1 | $5600^{+5700}_{-3500}$ ($6700^{+3000}_{-2500}$) | $3.1^{+1.5}_{-1.2}$ | $2.2^{+2.2}_{-1.6}$ |
|  | 2 | $5500^{+5200}_{-3100}$ ($5800^{+1300}_{-1200}$) | $3.1^{+1.5}_{-1.1}$ | $1.8^{+1.3}_{-1.3}$ |
|  | 4 | $4900^{+3100}_{-2300}$ ($5430^{+670}_{-640}$) | $2.9^{+1.6}_{-1.0}$ | $1.4^{+0.8}_{-1.0}$ |
|  | 6 | $5300^{+2300}_{-2200}$ ($5240^{+460}_{-440}$) | $2.8^{+1.4}_{-0.8}$ | $1.4^{+0.8}_{-0.9}$ |
| S2COSMOS | – | $5400^{+1600}_{-2200}$ | $3.1^{+1.1}_{-0.8}$ | $1.6^{+0.6}_{-0.6}$ |
| Cumulative | 1 | $4900^{+6000}_{-3500}$ ($6900^{+2500}_{-2200}$) | $3.2^{+1.9}_{-1.2}$ | $2.1^{+1.9}_{-1.5}$ |
|  | 2 | $4800^{+4900}_{-3000}$ ($5820^{+1100}_{-990}$) | $3.0^{+2.0}_{-1.1}$ | $1.7^{+1.1}_{-1.3}$ |
|  | 4 | $4400^{+2400}_{-2000}$ ($5510^{+560}_{-530}$) | $2.9^{+1.6}_{-0.9}$ | $1.3^{+0.9}_{-1.0}$ |
|  | 6 | $5000^{+1700}_{-1800}$ ($5220^{+350}_{-360}$) | $2.6^{+1.2}_{-0.7}$ | $1.2^{+0.8}_{-0.9}$ |
| S2COSMOS | – | $5100^{+1000}_{-1500}$ | $3.0^{+0.8}_{-0.6}$ | $1.5^{+0.5}_{-0.6}$ |

### 4.2 The environments of radio AGN in COSMOS

When selecting the RQ sample we excluded any galaxies that had been flagged as 'MLAGN' or 'HLAGN' in the VLA-COSMOS catalogue (Smolčić et al. 2017b; blue triangles in Fig. 1), so as to minimize any contamination by radio-loud sources (Section 2.1). We are therefore able to repeat the construction of the number counts, but in environments around galaxies with AGN-driven radio emission. This sample consists of 148 galaxies, with a median of six matched in mass and redshift to each of the 18 considered RAGERS H$z$RGs.

Note that $< 1.5$ per cent of these 148 VLA-COSMOS MLAGN/HLAGN analogues to RAGERS H$z$RGs are 'radio-loud': applying the same method as described in Section 2.1 to estimate their rest-frame 500 MHz luminosities, we found that only two of the 148 MLAGN/HLAGN (1.35 per cent) have $L_{500\,\mathrm{MHz}} > 10^{27}$ W Hz$^{-1}$ (the traditional definition for 'radio-loud' galaxies and H$z$RGs, and the cutoff for RAGERS galaxies; Miley & De Breuck 2008), with the median of the distribution being $\log(L_{500\,\mathrm{MHz}}/\mathrm{W\,Hz^{-1}}) = 24.2^{+0.7}_{-0.5}$. The majority (98.6 per cent) of these galaxies are therefore not true H$z$RGs and have significantly fainter radio luminosities than the RAGERS sample. We emphasize that while the median luminosity for the MLAGN/HLAGN sample is similar to that quoted for galaxies in our RQ sample ($L_{500\mathrm{MHz}} = 10^{24.1\pm0.5}$ W Hz$^{-1}$; see Section 2.1), the latter was calculated using the 14.6 per cent of RQ analogues with counterparts in the VLA-COSMOS catalogue, such that the true distribution of $L_{500\,\mathrm{MHz}}$ for our RQ analogues likely extends to much lower values. The MLAGN/HLAGN sample thus probes an intermediate regime between H$z$RGs and our RQ sample.

We repeat the construction of the number counts and Schechter function fitting within 1, 2, 4, and 6 arcmin of galaxies in this MLAGN/HLAGN sample and find no significant difference with respect to either the blank field or the RQ environment number counts. Whilst in contrast with studies of SMGs around H$z$RGs (e.g. Ivison et al. 2000; Stevens et al. 2003, 2010; Greve et al. 2007; Rigby et al. 2014) this result is likely due to the MLAGN/HLAGN sample not being traditional radio-loud galaxies, and instead having radio emission that is more similar to RQ galaxies.

### 4.3 Sensitivity to overdensities

In order to interpret the significance of the apparent similarity between the environments of the RQ analogues and the blank field the next step is to determine the strength of overdensity that is required for a signal to be detected using our analyses. To address this question we measure the counts from randomly drawn samples of mock submillimetre sources, increasing the sample size (i.e. equivalent of $N_0$) to find the minimum number of sources required to measure number counts that are significantly different to those of the blank field.

The procedure is as follows, and is repeated for each of the four spatial scales studied (1, 2, 4, and 6 arcmin). First, each bin centre (or lower bin edge in the case of cumulative counts) is assigned a probability of selection based on the shape of the best-fitting blank-field Schechter function (Section 4.1). An initial number of simulated galaxies, $N_{\mathrm{gal}}$, is generated based on these probabilities and each simulated galaxy is assigned flux density uncertainties that match the median values of real S2COSMOS sources in the relevant flux bin. As with the calculation of the real number counts (Section 3) we then create $10^4$ realizations of the simulated catalogue and the entire process – from randomly choosing $N_{\mathrm{gal}}$ flux densities onwards – is repeated 1000 times for each $N_{\mathrm{gal}}$, such that each bin has a distribution of $10^7$ possible counts associated with it. The number counts in each bin is then the medians of these values, and the uncertainties account for both Poissonian uncertainties and the 16th–84th percentile ranges.

A Schechter function of the form described by equation (1) (or its integral described by equation 2 for the cumulative counts) is fitted to the resultant number counts by fixing $S_0$ and $\gamma$ to the blank-field values and scaling $N_0$, as was done for the real RQ galaxies (Section 4.1). We then define the quantity $\phi$, to parametrize the relative measured density of the simulated number counts, such that:

$$\phi = \frac{N_0^{\mathrm{fit}}}{N_0^{\mathrm{bf}}} - 1, \quad (3)$$

where $N_0^{\mathrm{fit}}$ is the best-fitting value to the simulated number counts and $N_0^{\mathrm{bf}}$ is the blank-field value. Thus, $\phi = 0$ indicates number counts that are identical to those of the blank field. The significance of an overdensity in the simulated number counts is given by the ratio of $\phi$ to its $1\sigma$ uncertainty. If this ratio is greater than unity, then the overdensity has a significance of $> 1\sigma$. This procedure is repeated until the value of $N_{\mathrm{gal}}$ converges on the minimum number of galaxies required for a $1\sigma$ overdensity to be detected, which is parametrized as $N_{\mathrm{gal}}^{\mathrm{min}}$.

To translate $N_{\mathrm{gal}}^{\mathrm{min}}$ into terminology more commonly used in protocluster studies, we calculate the overdensity parameter, $\delta$, which for a given data set is defined as:

$$\delta = \frac{N_{\mathrm{gal}}^{\mathrm{data}}}{N_{\mathrm{gal}}^{\mathrm{bf}}} - 1, \quad (4)$$

where $N_{\mathrm{gal}}^{\mathrm{data}}$ is the number of galaxies in the data and $N_{\mathrm{gal}}^{\mathrm{bf}}$ is the number of galaxies in the blank field across the same flux density range and area as the data set. Environments for which $\delta > 0$ are therefore overdense relative to the blank field, while those with $\delta < 0$ are underdense. In our estimate of the minimum overdensity to which our method is sensitive, $N_{\mathrm{gal}}^{\mathrm{data}}$ is substituted for $N_{\mathrm{gal}}^{\mathrm{min}}$, while the calculation of $N_{\mathrm{gal}}^{\mathrm{bf}}$ depends on the type of number counts (differential or cumulative): for differential number counts, $N_{\mathrm{gal}}^{\mathrm{bf}}$ is calculated by summing all bins $> 3$ mJy after multiplying by the simulated area and the bin widths; for cumulative number counts it is given by the value of the 3 mJy bin multiplied by the simulated area.

This analysis shows that our study of differential submillimetre number counts is sensitive to overdensities with $\delta \gtrsim$ 1.2, 0.93, 0.86, and 0.85 for radii of 1, 2, 4, and 6 arcmin, respectively. For the cumulative counts, we are sensitive to $\delta \gtrsim$ 0.47, 0.40, 0.38, and 0.37 for 1, 2, 4, and 6 arcmin radii. Thus, the lack of detections in any of our samples suggests that RQ galaxies are in regions with $\delta \lesssim 0.4$ at submillimetre wavelengths. For comparison, Rigby et al. (2014) found values of $\delta$ ranging from $-0.27$–0.9 for 500 µm-selected sources in known protoclusters around H$z$RGs at $z \sim$ 2–4, using a search radius of 3.5 arcmin. Targeted 870 µm observations of the ~140 arcmin$^2$ region around the H$z$RG MRC1138−262 at $z = 2.16$ revealed an overdensity of SMGs with $\delta \sim$ 1–3 (Dannerbauer et al. 2014). Meanwhile the 1.1 mm number counts presented by Zeballos et al. (2018) indicate SMG overdensities of $\delta \sim 1$ in 3/16 of their target H$z$RG fields, and an overdensity of $\delta \gtrsim 2$ when all of their target fields are combined and the central 1.5 arcmin regions around the central AGN considered. Thus, overdensities commensurate with those around H$z$RGs would have been detected by our study of submillimetre number counts around RQ galaxies, and the absence of the detection of a significant overdensity requires that RQ galaxies at $z = 1$–3 are in less overdense environments than H$z$RGs of similar masses. The implications of this finding are discussed further in Section 5.

### 4.4 Density of faint sources

Single-dish submillimetre surveys (including S2COSMOS) are affected by confusion and high backgrounds: much of the 'noise' in the maps is from a background of faint sources. By studying the statistic of noise peaks in the maps we can therefore probe the distribution of galaxies that are below the flux limit of the catalogue (e.g. Glenn et al. 2010; Viero et al. 2013).

We next investigate the environments of the RQ analogues using the S2COSMOS signal-to-noise (SNR) map to track whether there is an overdensity of faint submillimetre sources in these regions. SNR peaks are identified in the map using the PYTHON package PHOTUTILS (Bradley et al. 2022) with detection thresholds from 1.5–4 and the surface density of these peaks within 1, 2, 4, and 6 arcmin of each RQ analogue is calculated. The blank-field density is estimated (for each aperture radius) by randomly placing $10^4$ apertures across the SNR map and repeating the calculation.

We compare the blank-field and RQ galaxy environments by comparing the surface density distributions between the regions around RQ galaxies and the blank field, as shown in Fig. 4 for the SNR > 1.5 detections (top four panels) and the SNR > 4.0 detections (bottom four panels). To quantitatively compare the statistics of SNR peaks in the blank field and near RQ galaxies we perform a two-sample Kolmogorov–Smirnov (KS) test on the resultant distributions.

For SNR detection thresholds < 2.5 (e.g. top four panels of Fig. 4) the KS test shows that there is no significant difference between the blank-field distributions at any radii ($p > 0.11$, and typically $p > 0.4$). For the higher SNR thresholds (e.g. bottom four panels of Fig. 4), where a larger fraction of the SNR peaks are likely real sources, the $p$-values exhibit a trend such that at the largest radii the distributions of source density between the blank-field and RQ galaxies are likely drawn from the same distribution, but they start to show hints of different distributions at the smallest radii. The most significant result is in the 1 arcmin radii search for SNR > 4 peaks in the map (i.e. the closest analogue to using the S2COSMOS catalogue directly), where $p = 7 \times 10^{-6}$, corresponding to a ~4.5$\sigma$ confidence that the distribution between RQ environments and the blank field are different at this scale and SNR limit. The KS test significance drops to ~2.5$\sigma$ at 2 arcmin ($p = 0.026$) and is even lower at larger radii and smaller SNR thresholds. If this result is confirmed then it suggests that RQ galaxies may be marginally overdense at small scales (~ 1 arcmin, which corresponds to ~ 0.5 Mpc at $z \sim 2$) for faint sources at submillimetre wavelengths, when compared to the blank field. We caution however that these differences could be driven by small number statistics, as there are few objects within these small radii.

We perform the same analysis for the sample of HLAGN/MLAGN galaxies discussed in Section 4.2 but find no significant difference compared with the blank field, obtaining $p > 0.34$ regardless of search radius or SNR threshold. The only possible exception occurs when the SNR threshold is set to > 1.5 and a search radius of 1 arcmin is used; in this instance $p = 0.004$, corresponding to ~ 2.9$\sigma$ significance. However, given the small size of the HLAGN/MLAGN sample (148 galaxies; Section 4.2) and the rapid decline in the significance of the discrepancy as the SNR threshold increases (dropping to < 1.8$\sigma$ for SNR > 1.6, and to < 1.4$\sigma$ for SNR > 1.7), we attribute this signal as an anomaly driven by small number statistics.

### 4.5 The environments of individual galaxies

The results presented thus far quantify the SMG density in the average environments of high-mass RQ galaxies, and in HLAGN/MLAGN of similar masses and redshifts, all of which are selected as analogues to the RAGERS H$z$RG sample (see Section 2.1). While this is our primary goal, one question that we can also address is whether any individual galaxies in our samples are seen to reside in significant SMG overdensities.

We thus measure the SMG overdensity in the environment of each RQ/HLAGN/MLAGN galaxy using the following procedure: First, any sources from the S2COSMOS catalogue that lie within $R = 1$, 2, 4 and 6 arcmin of the target galaxy are identified, and their flux densities and corresponding completeness corrections retrieved from each of the $10^4$ realizations of the catalogue created in Section 3. For each realization, we exclude any sources whose flux density is < 3 mJy and count the remaining sources within the desired separation $R$ from the target galaxy, applying a completeness correction for each source. The median of the resultant distribution is then taken to represent $N_{\rm gal}^{\rm data}$ from equation (4), and uncertainties are estimated using the 16th–84th percentile range and factoring in Poissonian uncertainties. The overdensity $\delta$ is then calculated via equation (4), where $N_{\rm gal}^{\rm bf}$ is estimated by multiplying the 3 mJy bin from the blank-field cumulative number counts by the area being probed (i.e. $\pi R^2$ in deg$^2$).

For each environment we quantify the significance of the overdensity or underdensity as the ratio of its value to its uncertainty, i.e. $\delta/\sigma_\delta$. Since the uncertainties on $\delta$ are typically asymmetric, different treatment is required for overdensities and underdensities: for overdensities (i.e. $\delta > 0$) we use the uncertainty in the negative direction as $\sigma_\delta$, while for underdensities (i.e. $\delta < 0$) we use the uncertainty in the positive direction. Environments for which $|\delta|/\sigma_\delta \leq n$ are then considered consistent with the blank field to within $n\sigma$.

Fig. 5 shows the distributions of $\delta/\sigma_\delta$ for our RQ and HLAGN/MLAGN samples. Note that we opt to show this quantity rather than $\delta$ itself in order to incorporate the variation in uncertainty from system to system. Regardless of the search radius used ($R = 1$, 2, 4 or 6 arcmin), the majority of galaxies lie within the region bounded by $\delta/\sigma_\delta = \pm 1$ (the shaded, hatched region in Fig. 5) and are thus consistent with the blank field to within $1\sigma$. For comparison, we also plot on Fig. 5 a Gaussian distribution with a mean and variance of zero and unity, respectively. For all but the 1 arcmin search radius there is no evidence of deviation from the Gaussian distribution in both the RQ and HLAGN/MLAGN samples. In the case of the 1 arcmin search radius (top left panels in Fig. 5) a Gaussian distribution is a poor match to the observations. However, in this case, underdensities are hard to identify due to the small number of galaxies in the search area. This dearth of underdense regions is the likely cause of the observed non-Gaussianity. Excluding this case, we thus find no excess of overdensities relative to Gaussian expectations, implying that there is no systematic tendency for either sample to reside in significantly overdense regions. These findings are consistent with the number counts for each sample presented in Section 4.1 and Section 4.2.

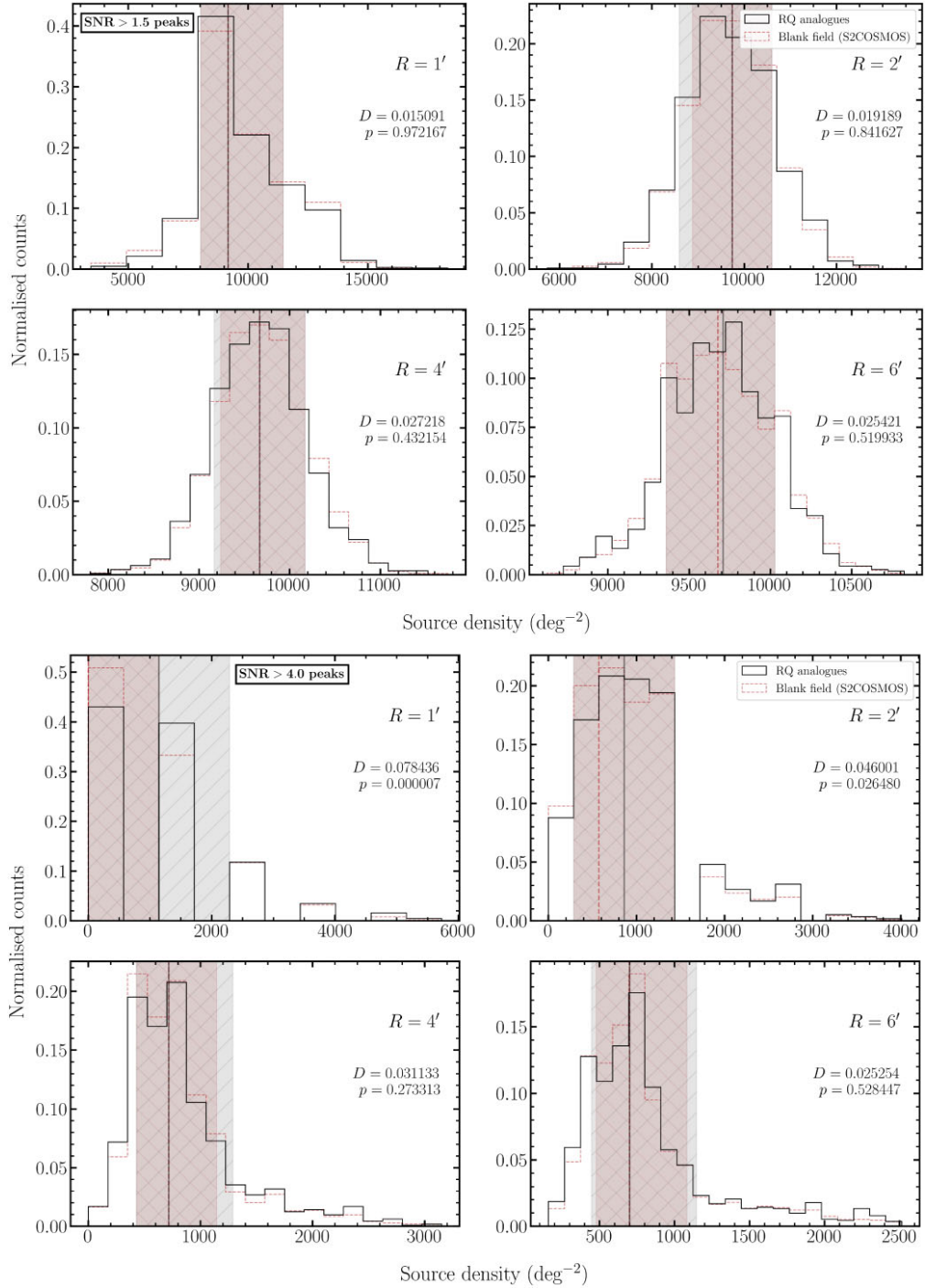

**Figure 4.** Distribution of the density of submillimetre peaks with SNR > 1.5 *(top four panels)* and SNR > 4 *(bottom four panels)* around RQ galaxies compared with the blank field. The histograms compare regions of radius, $R = 1, 2, 4,$ and 6 arcmin (as labelled) around RQ galaxies with the equivalent blank field area. Vertical lines and shaded regions show the median and 16th–84th percentiles of the distributions, which significantly overlap in most panels. Results from two-sample KS-tests ($p$ and $D$) are printed on the right of each panel and show that for the SNR > 1.5 peaks *(top)* in all four cases the regions around RQ galaxies are consistent with being drawn from the same distribution as the blank field. As discussed in Section 4.4 the results are similar for all radii and SNR thresholds tested, though at the bright end there is a hint of overdensity at the smallest scales as shown by the SNR > 4 histograms *(bottom)*.

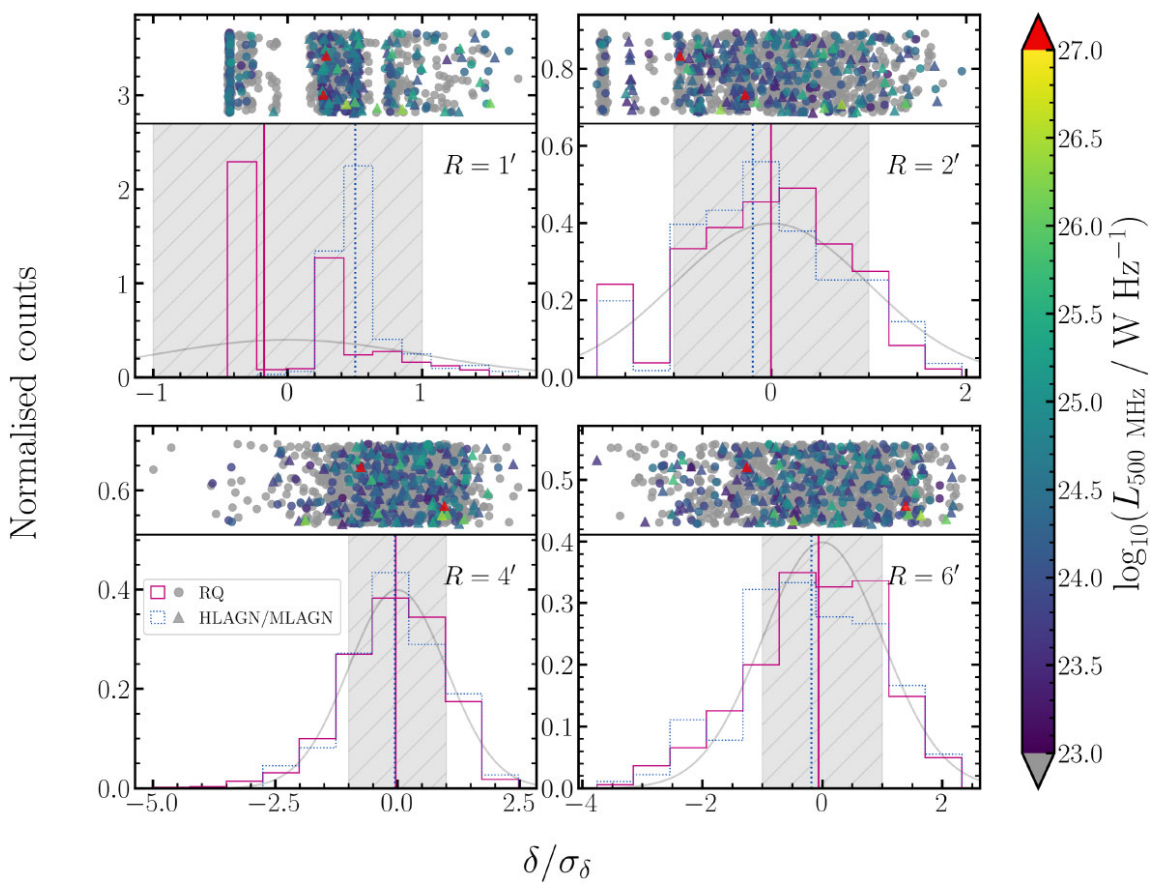

**Figure 5.** Distributions of the ratio of the SMG overdensity parameter $\delta$ to its uncertainty $\sigma_\delta$, which we use as a proxy for the significance of a given overdensity/underdensity in the environments of galaxies in our RQ (magenta, solid) and HLAGN/MLAGN (blue, dotted) samples. Each panel is labelled with the radius used to search for SMG companions. At all radii we find that the majority of galaxies in both samples are consistent with the blank field to within $1\sigma$, as indicated by the shaded and hatched region between $\delta/\sigma_\delta = \pm 1$. Furthermore the distributions are generally centred around $\delta/\sigma_\delta \sim 0$, and are roughly symmetric except when the smallest radius is used, at which point small number statistics are expected to make the detection of underdensities difficult. The grey curve in each panel shows a Gaussian distribution with a mean of zero and a variance of unity, which approximately matches the observed distributions. Circles (triangles) at the top of each panel show the value of $\delta/\sigma_\delta$ for each individual galaxy in the RQ (HLAGN/MLAGN) sample, with randomly chosen positions along the *y*-axis. These points are coloured according to the rest-frame 500 MHz luminosity of the galaxy ($L_{500\text{MHz}}$; see Section 2.1 for details) with VLA-COSMOS non-detections shown in grey. Overall we see no trend between $L_{500\text{MHz}}$ and $\delta/\sigma_\delta$.

Since many of the galaxies in these samples have been detected in VLA-COSMOS (14.6 per cent of the RQ galaxies and 100 per cent of the HLAGN/MLAGN; Smolčić et al. 2017b), we also investigate any potential dependence of $\delta/\sigma_\delta$ on the rest-frame 500 MHz radio luminosities calculated in Section 2.1. To this end we add circles (triangles) to the top of each panel in Fig. 5 to show the values of $\delta/\sigma_\delta$ for the environments of individual galaxies in the RQ (HLAGN/MLAGN) sample, with positions along the *y*-axis chosen randomly for visualization purposes, and colours according to their rest-frame 500 MHz luminosities (VLA-COSMOS non-detections are shown in grey). We see no overall correlation between the radio luminosity of a galaxy and the value of $\delta/\sigma_\delta$ for its environment. Of particular interest are the two HLAGN/MLAGN galaxies with $L_{500\text{MHz}} > 10^{27}$ W Hz$^{-1}$ (red triangles in Fig. 5), as these would meet the criterion for being HzRGs (Miley & De Breuck 2008). One might then expect them to reside in SMG overdensities such as those identified around other HzRGs (e.g. Ivison et al. 2000; Stevens et al. 2003, 2010; Greve et al. 2007; Dannerbauer et al. 2014; Rigby et al. 2014; Zeballos et al. 2018), yet they show no signs of residing in significantly overdense environments regardless of the radius used to identify possible SMG companions. The only possible exception occurs when a radius of 6 arcmin is used, at which point one of the two galaxies resides in a $\sim 1.4\sigma$ overdensity. However, even then the other is located in a $\sim 1.3\sigma$ underdensity on these scales, such that on average there is no tendency for these galaxies to reside in SMG overdensities. This would however be consistent with the significant field-to-field variation seen for HzRGs at 500 µm (Rigby et al. 2014) and at 1.1 mm (Zeballos et al. 2018). Future comparison of these galaxy environments with those of the stellar mass- and redshift-matched RAGERS HzRGs will help in understanding the cause of this variation.

## 5 DISCUSSION

In this study, we used number counts to show that massive RQ galaxies at $z = 1$–3 reside in regions that have similar submillimetre source density to blank-field regions. The RQ galaxies have $\delta \lesssim 0.4$, though our constraints are marginally stronger on larger scales (up to $\sim 3$ Mpc) and weaker on smaller scales (down to to $\sim 0.5$ Mpc;

Section 4.3). Similarly, our study of peaks in the 850 µm SNR map found that the regions around massive RQ galaxies are mostly consistent with being drawn from the same distribution as blank-field regions, although there is a hint of some overdensity on < 1 arcmin (∼ 0.5 Mpc) scales.

The sample of RQ galaxies analysed was selected to match specific H$z$RGs in stellar mass and redshift (Section 2), the nature of whose environments is as yet unknown. We therefore emphasize that the goal of this paper is to pave the way for a comparison of the submillimetre environments of RQ galaxies and H$z$RGs, controlled for stellar mass and redshift; this will be conducted in a future RAGERS paper (Greve et al. in prep.). Even so, there are several known examples of H$z$RGs residing in regions that are overdense in the submillimetre (e.g. Ivison et al. 2000; Stevens et al. 2003, 2010; Greve et al. 2007; Dannerbauer et al. 2014; Rigby et al. 2014; Zeballos et al. 2018), albeit with significant field-to-field variation seen at both 500 µm (Rigby et al. 2014) and at 1.1 mm (Zeballos et al. 2018). The lack of SMG overdensity seen in the environments of our RQ galaxies therefore suggests that there is difference between the submillimetre environments around massive RQ galaxies and H$z$RGs at $z \sim 1$–3. This implies that either the AGN or the radio emission has direct impact on the environment and star-formation activity in galaxies around H$z$RGs, or the H$z$RGs themselves are preferentially located in overdense regions, including regions with a lot of star formation in submillimetre sources.

We also investigated the number counts in regions around galaxies with radio emission and classified as MLAGN or HLAGN by Smolčić et al. (2017b) and find no significant overdensities around these galaxies. These radio galaxies have significantly lower radio luminosity than 'classic' H$z$RGs, and this non-detection also suggests that they reside in environments for which $\delta \lesssim 0.4$. Combined with our findings for the RQ sample, this implies that the density of the surrounding environment is not linked simply to the presence of an AGN; only when these AGN are radio-loud is there a preference towards residing in overdensities.

Overall, our results are consistent with a picture similar to those discussed by Wylezalek et al. (2013), in which regions of higher galaxy density impact the production of jets and radio emission from AGN. For example, galaxy mergers in overdensities may increase the spin of black holes, which makes them more able to power radio jets (e.g. Wilson & Colbert 1995; Sikora, Stawarz & Lasota 2007). Another possibility is the jet confinement theory, which proposes that radio synchrotron emission may be brightened by interaction with a denser intergalactic medium (IGM; Barthel & Arnaud 1996). Our study of the SNR peaks in the SCUBA-2 map suggested that RQ galaxies may be in small overdensities on $\lesssim 0.5$ Mpc scales (Section 4.4). If this result is found to be robust (e.g. in studies of larger samples, or deeper data) then it cannot be caused by interaction of radio jets with the IGM (since large-scale jets are not present in RQ galaxies). Instead such a result would suggest that some of the observed overdensity around these RQ galaxies and their RL counterparts is due to their high stellar masses predisposing them to occupy high-density environments, but with H$z$RGs being most likely to be present in the most massive haloes due to galaxy mergers or interaction with the IGM.

It is intriguing to note that our results and these hypotheses are in contrast with an initial examination of the SHARK semi-analytic model of galaxy formation (Lagos et al. 2018). Detailed analyses of the submillimetre environments of simulated H$z$RGs and otherwise similar RQ galaxies in SHARK will be presented in Vijayan et al. (in prep.). The forthcoming analyses of our new SCUBA-2 observations of the mass- and redshift-matched H$z$RG sample in RAGERS will enable confirmation of our detection of a difference in the submillimetre environments of H$z$RGs and RQ galaxies (Greve et al. in prep.).

We caution that while we have not detected significant overdensities of SMGs in the environments of massive RQ galaxies (and of similarly massive galaxies classified as HLAGN/MLAGN by Smolčić et al. 2017b), this does not mean that these galaxies do not reside in overdense environments. First we are limited by the lack of redshift information for the submillimetre sources in S2COSMOS, consequently having to restrict our search to projected overdensities; it is possible that many of the galaxies examined in this study reside in physical (volumetric) overdensities which have been smoothed out by the inclusion of foreground/background galaxies in our study. Secondly, searches in the submillimetre regime are inherently biased towards gas-rich galaxies with high star formation rates, such that galaxies undergoing less active star formation will likely lie below the confusion limit of S2COSMOS and thus evade detection. Indeed, one might expect galaxies in the environments of H$z$RGs to be inherently more gas rich than those near RQ galaxies: several studies have indicated that star formation activity approximately scales with AGN activity (e.g. Florez et al. 2020; Zhuang & Ho 2020; Xie et al. 2021; Zhuang, Ho & Shangguan 2021), likely as a result of both being fuelled by similar reservoirs of gas in the host galaxy (e.g. Sanders et al. 1988; Hopkins et al. 2008; Zhuang et al. 2021). Galaxies in the same large-scale structure as a radio-loud (and therefore likely gas-rich) AGN host may then also be similarly gas rich, owing both to their common heritage and to their shared environment. Conversely, RQ galaxies with little to no nuclear activity may tend to reside in gas-poor environments, surrounded by galaxies that are similarly gas poor and thus difficult to detect in the submillimetre regime. We therefore cannot altogether rule out the possibility that these RQ galaxies reside in overdensities of gas-poor galaxies.

## 6 CONCLUSIONS

We have conducted a search for 850 µm-selected SMGs in the environments of massive, radio-quiet galaxies at $z \sim 1$–3 in the COSMOS field. The sample of RQ galaxies was selected to match the stellar masses and redshifts of H$z$RGs so our results can be compared with studies of H$z$RGs. Our main conclusions are as follows:

(i) Using data from the S2COSMOS catalogue (Simpson et al. 2019) we constructed number counts in the regions of the RQ galaxies and compared these with the blank field to determine whether massive, $z \sim 1$–3 RQ galaxies typically reside in overdense regions, as is expected of their radio-loud counterparts. No significant difference is detected between the number counts for the environments of the RQ galaxies and those for the blank field. This result remains when examining regions from 1 to 6 arcmin scales and using both differential and cumulative number counts. It also holds both for completely free Schechter function fits and when fixing $S_0$ and $\gamma$ to the blank-field values to pinpoint any difference in $N_0$.

(ii) We tested the sensitivity of our analyses to identifying SMG overdensities by using simulated number counts, and found that we can detect overdensities with $\delta \gtrsim 0.4$. This threshold is sufficient to identify many known protoclusters, though there is significant variation between fields, particularly at submillimetre wavelengths.

(iii) A similar examination of the submillimetre number counts around galaxies detected in the radio and classified MLAGN or HLAGN by Smolčić et al. (2017b) found that these sources are also in environments that are statistically indistinguishable (i.e. $\delta \lesssim 0.4$

using our method) from the blank field. These galaxies have some radio emission, but they are not H$z$RGs and have median rest-frame 500 MHz luminosity that is nearly three orders of magnitude fainter than H$z$RGs.

(iv) To probe faint sources not individually detected in the S2COSMOS catalogue we also investigated the distribution of SNR peaks in the 850 µm map and used KS tests to search for differences between the region around massive, $z \sim$ 1–3 RQ galaxies and the blank field. We test detection thresholds of SNR > 1.5 up to SNR > 4 (similar to the S2COSMOS catalogue) and regions of 1–6 arcmin radius, finding that the density of submillimetre peaks around RQ galaxies is consistent with the blank field in most cases. For the higher SNR cuts and the smaller radii the KS test $p$ statistic is smallest, and suggests that there may be some weak overdensities around RQ galaxies when compared to the field.

(v) We calculated the overdensity parameter $\delta$ for individual galaxies in both the RQ and HLAGN/MLAGN samples, along with the corresponding uncertainty $\sigma_\delta$. Using $\delta/\sigma_\delta$ as a measure of the significance of a given overdensity, we find that while some individual galaxies in each sample reside in overdensities of > 1$\sigma$, the numbers of such overdensities do not exceed expectations from a Gaussian distribution, regardless of the search radius used. This reinforces the conclusion that there is no systematic tendency for galaxies in either the RQ or HLAGN/MLAGN samples to reside in overdense environments.

Thus, our analyses suggest that massive RQ galaxies at high redshift do not typically reside in substantial SMG overdensities. This contrast with previous studies of H$z$RGs (e.g. Rigby et al. 2014) suggests that the mechanisms powering RL galaxies have some link with the wider environment. We purport that this link may be driven by galaxy mergers in overdense environments elevating the accretion rate (and/or the spin) of the central black hole to produce more powerful radio jets. An alternative explanation is that a denser IGM enhances the synchrotron radiation emitted by radio jets, such that radio galaxies in overdense environments appear more luminous than their counterparts in lower-density environments. Future RAGERS papers will compare these findings in detail with results for the RL sample (Greve et al. in prep.) and with expectations from simulations (Vijayan et al. in prep.) and further explore the role of environment in regulating AGN activity.

## ACKNOWLEDGEMENTS

TMC received support from the Science and Technology Facilities Council (STFC; 2287406) and the Faculty of Science and Technology at Lancaster University. JLW acknowledges support from an STFC Ernest Rutherford Fellowship (ST/P004784/2). The Cosmic Dawn Center (DAWN) is funded by the Danish National Research Foundation under grant No. 140. TRG is grateful for support from the Carlsberg Foundation via grant No. CF20-0534. LCH was supported by the National Science Foundation of China (11991052, 12233001), the National Key R&D Program of China (2022YFF0503401), and the China Manned Space Project (CMS-CSST-2021-A04, CMS-CSST-2021-A06). C-CC acknowledges support from the National Science and Technology Council of Taiwan (111-2112M-001-045-MY3), as well as Academia Sinica through the Career Development Award (AS-CDA-112-M02). X-JJ was supported by the National Science Foundation of China (12373026). DJBS acknowledges support from the UK Science and Technology Facilities Council (STFC) under grants ST/V000624/1 and ST/Y001028/1. HD acknowledges support from the Agencia Estatal de Investigación del Ministerio de Ciencia, Innovación y Universidades (MCIU/AEI) under grant (Construcción de cúmulos de galaxias en formación a través de la formación estelar oscurecida por el polvo) and the European Regional Development Fund (ERDF) with reference (PID2022-143243NB-I00/10.13039/501100011033).

The James Clerk Maxwell Telescope is operated by the East Asian Observatory on behalf of The National Astronomical Observatory of Japan; Academia Sinica Institute of Astronomy and Astrophysics; the Korea Astronomy and Space Science Institute; the National Astronomical Research Institute of Thailand; Center for Astronomical Mega-Science (as well as the National Key R&D Program of China with No. 2017YFA0402700). Additional funding support was provided by the Science and Technology Facilities Council of the United Kingdom and participating universities and organizations in the United Kingdom and Canada. Additional funds for the construction of SCUBA-2 were provided by the Canada Foundation for Innovation.

## DATA AVAILABILITY

The data underlying this article originate from three publicly available data sets. The COSMOS2020 data were previously published in Weaver et al. (2022), the S2COSMOS data in Simpson et al. (2019), and the VLA-COSMOS data in Smolčić et al. (2017b). They are available as described in these publications.

[1]*Department of Physics, Lancaster University, Lancaster LA1 4YB, UK*
[2]*Department of Physics, University of Oxford, Denys Wilkinson Building, Keble Road, Oxford OX1 3RH, UK*
[3]*DTU Space, Technical University of Denmark, Elektrovej 327, DK-2800 Kgs. Lyngby, Denmark*
[4]*Department of Physics and Atmospheric Science, Dalhousie University, Halifax B3H 4R2, NS, Canada*
[5]*Department of Physics and Astronomy, University of British Columbia, 6225 Agricultural Rd., Vancouver V6T 1Z1, Canada*
[6]*Academia Sinica Institute of Astronomy and Astrophysics (ASIAA), No. 1, Section 4, Roosevelt Road, Taipei 106216, Taiwan*
[7]*Instituto de Astrofísica de Canarias (IAC), E-38205 La Laguna, Tenerife, Spain*
[8]*Departamento Astrofísica, Universidad de la Laguna, E-38206 La Laguna, Tenerife, Spain*
[9]*Institute of Astronomy, National Tsing Hua University, No. 101, Section 2, Kuang-Fu Road, Hsinchu 30013, Taiwan R.O.C*
[10]*Kavli Institute for Astronomy and Astrophysics, Peking University, Beijing 100871, China*
[11]*Department of Astronomy, School of Physics, Peking University, Beijing 100871, China*
[12]*Research Center for Astronomical Computing, Zhejiang Laboratory, Hangzhou 311100, China*
[13]*East Asian Observatory, 660 North A'ohoku Place, Hilo, HI 96720, USA*
[14]*ARC Centre of Excellence for All Sky Astrophysics in 3 Dimensions (ASTRO 3D), Australia*
[15]*International Centre for Radio Astronomy Research, The University of Western Australia, 35 Stirling Hwy, 6009 Crawley, WA, Australia*
[16]*School of Physical Sciences, The Open University, Walton Hall, Milton Keynes MK7 6AA, UK*
[17]*Department of Earth Science Education, Kyungpook National University, 80 Daehak-ro, Buk-gu, Daegu 41566, Republic of Korea*
[18]*Centre for Astrophysics Research, University of Hertfordshire, College Lane, Hatfield AL10 9AB, UK*
[19]*Astronomy Centre, University of Sussex, Falmer, Brighton BN1 9QH, UK*
[20]*World Food Programme, No. 2 Jawatte Ave., Colombo 00500, Sri Lanka*

This paper has been typeset from a TEX/LATEX file prepared by the author.